\pdfoutput=1
\documentclass{JINST}
\usepackage[utf8]{inputenc}
\usepackage{amsmath}
\usepackage{graphicx}
\usepackage{url}
\usepackage[mathscr]{eucal}

\title{Macroscopic Dynamics of Entangled 3+1-Dimensional Systems: A Novel Investigation Into Why My MacBook Cable Tangles in My Backpack Every Single Day}
\author{ R. Dorrill$^a$, J. Felis$^b$\\ 
\llap{$^a$} Illinois Institute of Technology, \\ 10 West 35th Street Chicago, Il 60616, USA \\
\llap{$^b$} University of Hawaii at Meownoa, \\ 2500 Campus Rd, Honolulu, HI 96822, USA \\
}

\abstract{
A key axiom of equilibrium statistical physics is that all microstates are equally probably in a thermally isolated system. Coupled with the laws of Newtonian mechanics, quantum mechanics, chemistry, and thermal physics, one can build from this axiom both complex and satisfactory models for macroscopic phenomena. Here, we apply the precepts of statistical physics to a problem that has puzzled scientists and engineers since its discovery in the 1980's: The Entangled Laptop Cable Problem. Using a stochastic 2-dimensional simulation to approximate projections of the 3+1-dimensional system, we shall see that the overwhelmingly most probable state for a laptop cable is a severely tangled one.}

\keywords{High Energy Physics; Applied Physics; Statistical Physics; Machine Learning because why not}

\begin{document}

\maketitle
\section{Introduction and Motivation} \label{sec:intro}

The equilibrium behavior of mechanical systems is a well-known classical physics topic, generally considered to be understood in the present day. Such problems and their solutions are often assigned in undergraduate mechanics courses, usually for pedagogical purposes, though sometimes also as rites of passage. However, there remains a plethora of so-called low-hanging fruit - problems well known to the community but heretofore unsolved or plied with rigorous investigation. The Entangled Laptop Cable Problem is one such question. 

Here, we focus on a subset of the general problem: the entangled MacBook cable. The MacBook charging cable presents itself as a particularly interesting problem, for its phenomenology and equilibrium states are not well understood, and also because such devices are used by a large fraction of the field. These cables are also prone to tangling, in part due to the thin nature of the campure, and in part due to the MagSafe charging coupler, seen at the top of Fig. \ref{fig:RealCables}. While such cables formerly came with movable notches to wrap the cable around, that feature has been inexplicably removed, resulting in the monstrous topological entities seen in the figure.

If possible, it would be of great benefit to understand and solve this problem. The cost of wasted man-hours from disentangling such cables, when summed in aggregate across graduate students, postdocs, and professors at major R1 and teaching institutions, is surely astronomical. Preventing such tangles might not only save time and research efforts, but also protect the fragile sanity of faculty members when they finally arrive at work in the morning to find stale coffee, poor Rate My Professor reviews, and an absolute deluge of emails from students and administrators.

\begin{figure}[htb]
    \centering
    \includegraphics[trim = 0.0cm 0.0cm 0.0cm 0.0cm, clip=true, width=0.48\textwidth]{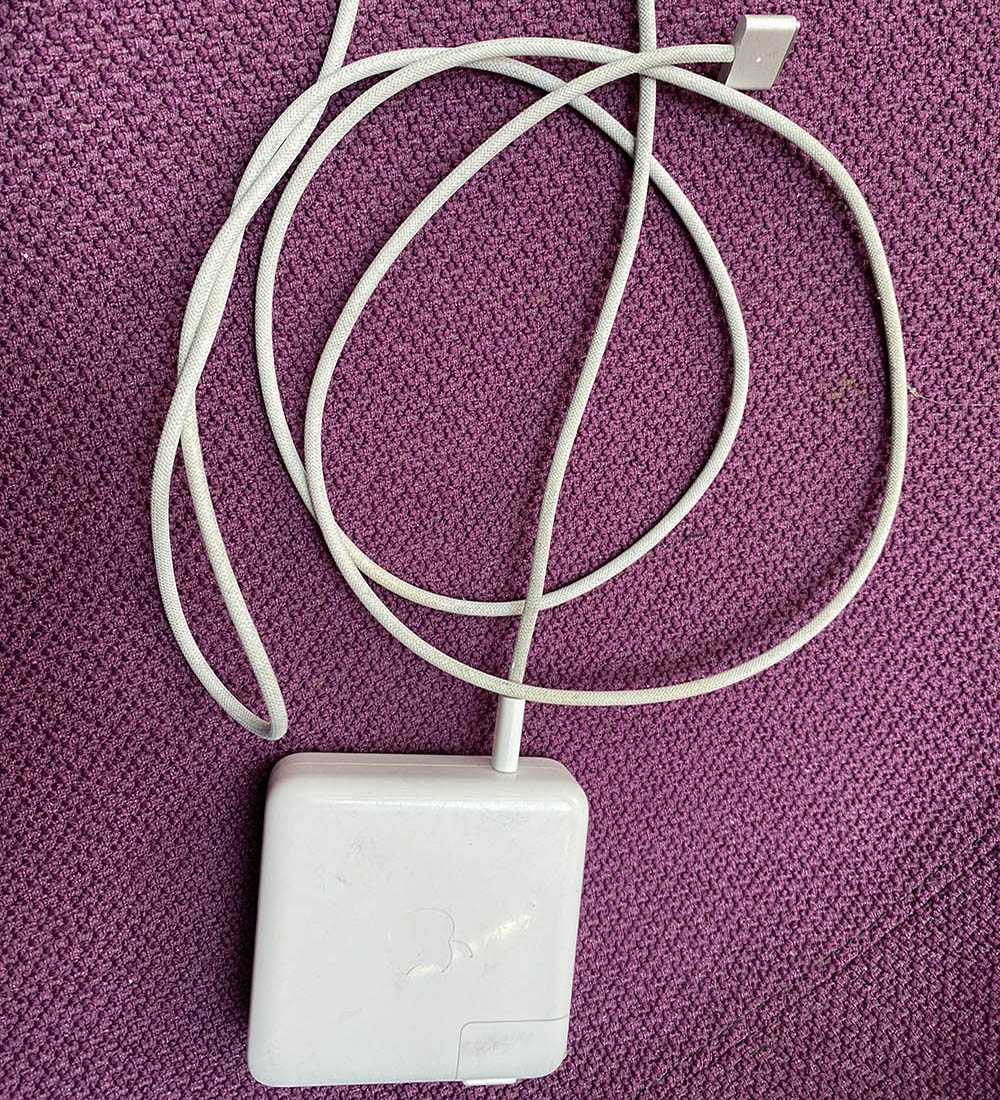} 
    \includegraphics[trim = 0.0cm 0.0cm 0.0cm 0.0cm, clip=true, width=0.48\textwidth]{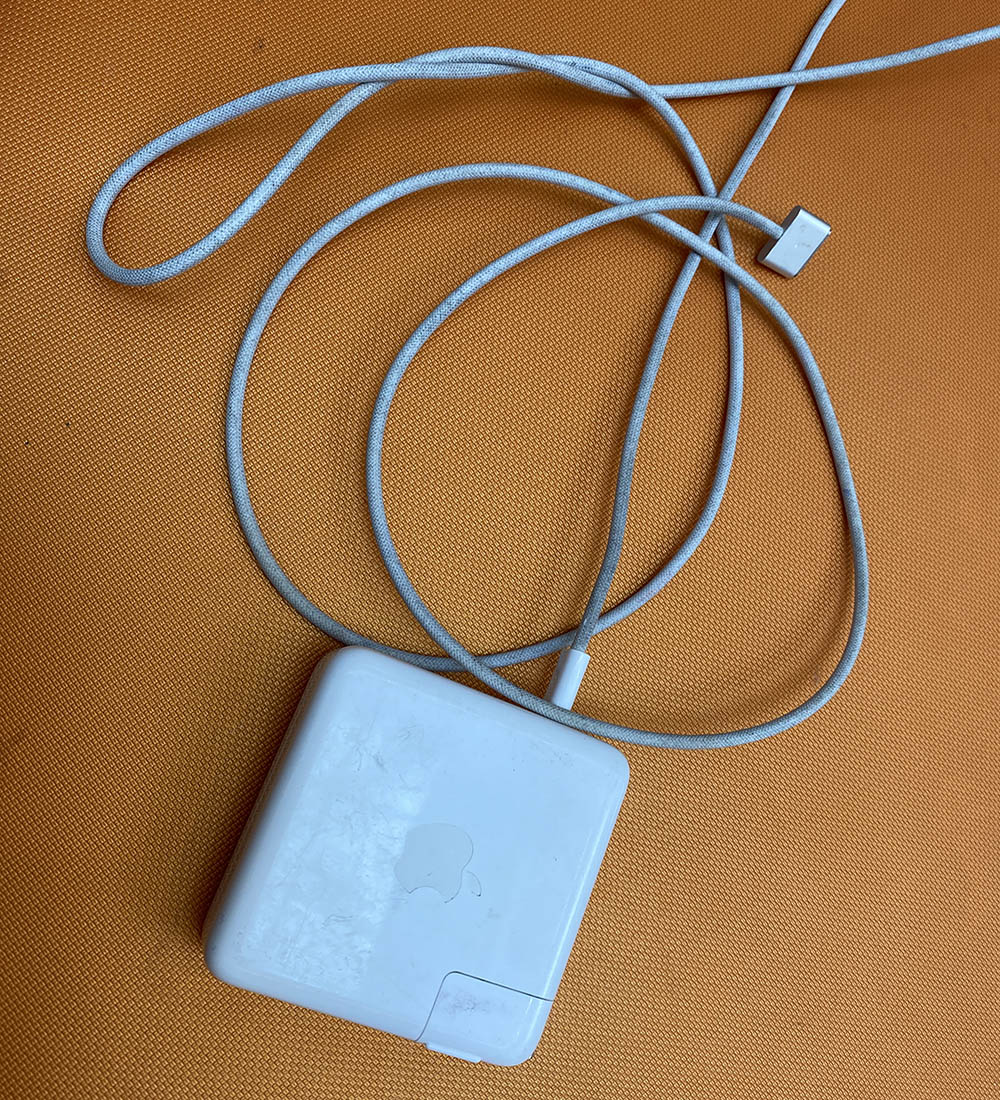} 
    \caption{Two tangled MacBook cable samples, collected in a sterile laboratory environment for further examination. Note the heavily twisted cable on the left, with a double helix shape, and the rectangular magnetic connector at the top of both photos. Such connectors are admittedly very useful, but also prone to snag on anything, with knots and snarls seemingly appearing out of nowhere.}
    \label{fig:RealCables}
\end{figure}

\section{Model and Theory} \label{sec:samples}

A mathematical representation for tangled laptop cables was sought to aid in understanding the problem. A plethora of potential functions seem worth consideration, and when considering the geometry of a cable one might first consider the edge cases:

\begin{itemize}
\item A long straight cable following a linear function in three dimensions: $ax+by+cz = C$. 
\item A circular or elliptical shape: $\frac{(x-h)^2}{a}+\frac{(y-k)^2}{b}= C$.
\item A neatly coiled helix, represented by a set of parametric equations: $x = rcos(t); y = rsin(t), z = ct$
\end{itemize}

In principal, the correct model would have some features of all of these distributions. Even most untangled cables will demonstrate some curvature, while tangled ones will have linear features, like those near the square adapter in the bottom of \ref{fig:RealCables}. For the purposes of the model, we define a class of non- novel functions here as \textit{tangles}, which are of the parametric form:\\
\[ x = x_{max}*cos(a{\pi}t)\]
\[y=(y_{max}*sin(b{\pi}t))*(c+dt+et^2+ft^3...)\]

\begin{figure}[htb]
\includegraphics[trim = 0.0cm 0.0cm 0.0cm 0.0cm, clip=true, width=0.99\textwidth]{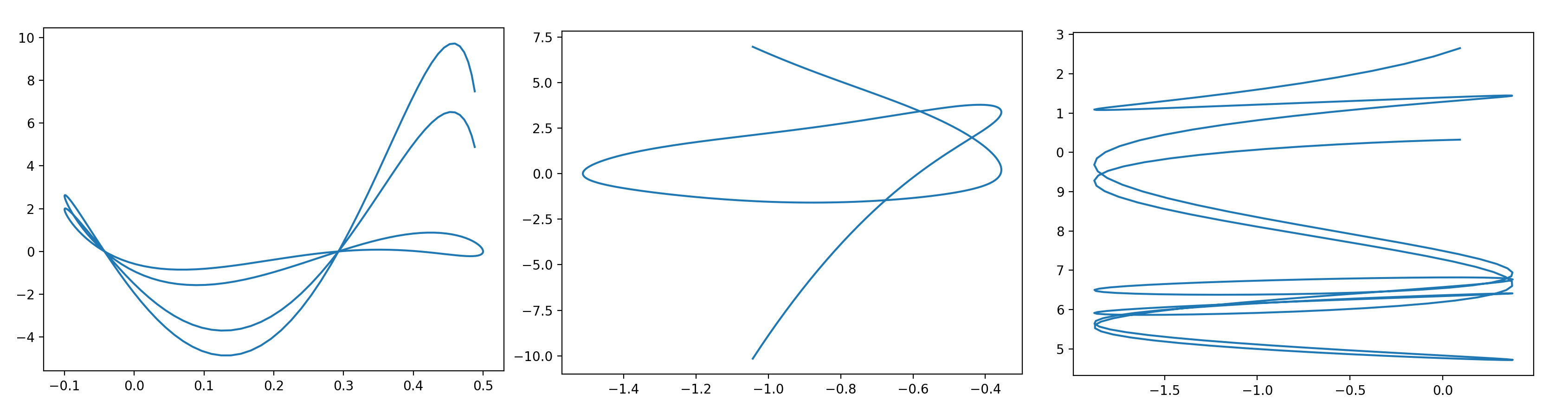} 
    \caption{Examples of 2-Dimensional projections of the three dimensional tangles. The parameters in each case were randomized, resulting in a standard tangled macbook cable (left), a short but nonetheless still-tangled cable (mid) , and an extra long, coil cable (right) . Descriptions of the model are provided in the text.}
    \label{fig:3tangles}
\end{figure}

Here, we omit the z-coordinate for simplicity, and frankly to save time because the authors are writing this at the last minute, although future investigations would do well to include z to account for the full complexity of the problem and the true nature of the mysterious 3+1-dimensional realm heretofore referred to as \textit{Backpack Space}, or BS. 

The tangle functions allow for a wide variety of behavior depending on the parameters, as seen in Fig. \ref{fig:3tangles}. While some appear to realistically represent the tangled monstrosities we regularly pull out of our backpacks, so as the left-hand or middle figure, other are overly trigonometric, or give a sense of unrealistic length or compactness. While such cases could also be interesting (see \textit{The Entangled Paracord Problem}) they are beyond the scope of this paper.

These functions have several tunable parameters of interest. In particular, we refer to 'b' and 'c' as the \textit{tangleosity}. Very low values (e.g. 0.001-0.5) result in low multiples of $\pi$ in the parametric arguments, and therefore low oscillatory behavior, low potential for tangling, and low \textit{crunodes}, i.e. locations where the function intersects with itself. These crunodes manifest in empirical tests as the knots and tangles in our MacBook cable. Note: the authors did not invent the term crunode, and mathematicians deserve the blame.\cite{enwiki:crunode} 

Low to Mid-range values of \textit{tangleosity} (1-10) result in fairly realistic approximations of tangled cables, particularly if one of the two parameters if kept on the low range, while the other is allowed to go higher. Finally, high values (10 or greater) result in chaotic geometries, particularly if both values are high, and unrealistic approximations of a real cable. Here, we attempt to define the realism, R, as the ratio of the length of the curve to the area of the plot, with units of 1/m. A higher R value means a much higher likelihood of tangling, but a much less realistic scenario. The cross-sectional area of a typical \textit{backpack space} gives a BS value of 0.135 (0.30x0.45m) while a typical MacBook cable is 2m in length, giving an R-factor of 14.8. The parameters $x_{max}$ and $y_{max}$ scale our curves, and give the cross-sectional area of our imaginary BS. 

To get some sense of the relative likelihood of tangled cables, we put some arbitrary limit on the parameters by selecting tangle coefficients from a normal distribution centered on 2$\pi$ with a standard deviation of 4, and polynomial coefficients for the y-parameters selected as special blend of random integers, with a minimum value of 2 (some parabolic behavior) and a max value of 11, with both the coefficients and number of coefficients randomized in such a way as to make this entire process convoluted and almost impossible to parse as we prepare our final results. The authors recommend a slightly simpler approach in followup papers.

\begin{table}[]
    \centering
    \begin{tabular}{|c|c|c|c|c|}
    \multicolumn{5}{c}{Example Functions}\\
    \hline
       \textbf{Tangleosities}  & \textbf{Poly Coefficients} & \textbf{Cable-like?} & \textbf{Crunodes} & \textbf{Degree of Tangle} \\
       \hline
       \hline
        [-6, 15] & [-1,7,3,4,-6,8] & Yes & 2 & Really Annoying\\
        \hline
        [ 4.90,14.92]& [0,11,-7,-13,-5,-11,14] & Yes & 2 & Not bad\\ 
        \hline
        [21.26,11.73] & [10,-4,11] & No & 8 & Day-ruining\\
        \hline
        [0.4,0.333]  & [-9,7,3,4,-6,8,9]  & Yes & 0 & None\\ 
        \hline
        [1.4$\pi$,1.6$\pi$]  & [-9,7,3,4,-6,8,9] & Yes & 3 & Meh \\
        \hline
        [6$pi$,6$\pi$]  & [-9,7,3,4,-6,8,9] & No & 4-5 & Argh \\
        \hline
        [6$\pi$,7$\pi$]  & [-9,7,3,4,-6,8,9] &  No & God only knows & Lovecraftian \\
        \hline
    \end{tabular}
    \caption{Properties of a sample of five \textit{tangles}.}
\end{table}

\begin{figure}[htb]
\includegraphics[trim = 0.0cm 0.0cm 0.0cm 0.0cm, clip=true, width=0.99\textwidth]{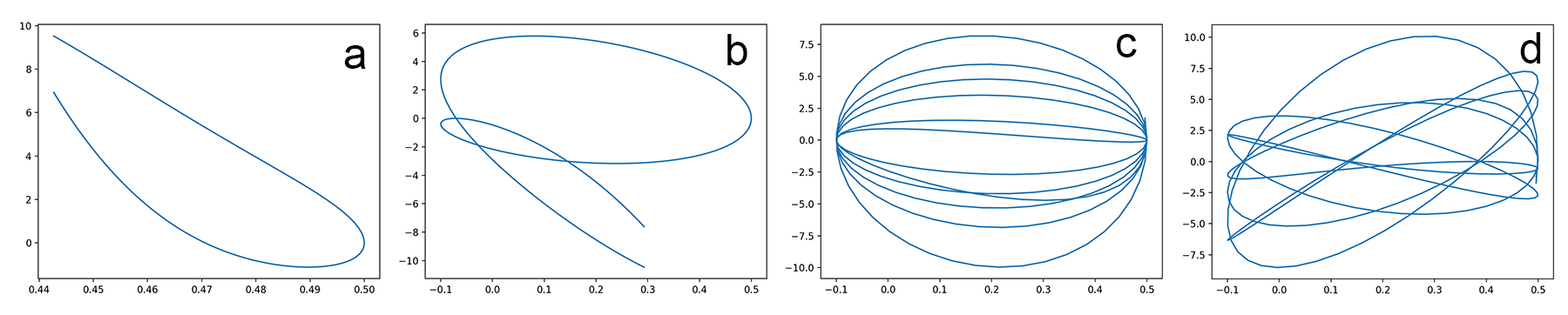} 
    \caption{Further examples of 2-Dimensional tangles. Here, for a particular set value of the polynomial coefficients, the tangleoisity was varied from low (a), to medium (b), and high (c and d). In tangle c, the x and y tangleosity were approximately equal, leading to a more elliptical shape. In (d), they different by a prime factor, leading to a chaotic geometry and many crunodes.}
    \label{fig:tangleositity}
\end{figure}

\section{Numerical Method and Strategy} \label{sec:exp}

With the \textit{Backpack Space} (BS) and tangle model in place, it was now possible to generate large samples of curves and rank them as tangled or untangled, based on their R-factor and number of crunodes. R-Factors below two, indicating a relatively low length, were countered as unrealistic, as were those above 35, although the latter were exceedingly rare. Admittedly, further work is needed to fine-tune these parameters and the randomized nature of the model to better resemble a true BS model. Large datasets of matched x-y coordinates were then generated for each randomized model, and visually plotted and scanned for crunodes using python's matplotlib functionality.\cite{Hunter:2007}

\section{Results and Discussion} \label{sec:results}

The resulting series of plots, scanned by hand for crunodes, or tangles, were ranked as "tangled" if they exhibited more than 3 crossings of the cable, which is judged as arbitraily likely to result in a tangle inside the backpack. Those with 3 or less crossings were judged as unlikely to be tangled. Here, with this initial model, we find a \textbf{$75.3\%$ likelihood of a tangled cable}, indicating that if one shoves their MacBook cable carelessly into their backpack, the overwhelmingly most likely state for the cable to end up in is the tangled one. This result, which confirming the experimental observations of BS tangles, nonetheless seems low given day-to-day experience.

We believe our model and numerical simulation needs further refinement, in part to match the arbitrary dimensions of the model functions to the real of actual BS, and perhaps to further automate the counting of crunodes and comparison of R-factors to allow for larger data-sets, less human error, and a more satisfactory result. Perhaps further studies might also offer a solution to the The Entangled Laptop Cable Problem, beyond the trivial one of taking 10 seconds to carefully pack the cable before work, or buying a better laptop bag to store the laptop, cable and accessories. We find such solutions unsatisfactory, as graduate students, postdocs, and professors are far too busy with committees to carefully pack their cables, and the most-populous lower rungs on the academic pyramid (i.e. grads and postdocs) are unlikely to have ample funding to invest in better hardware.


\begin{figure}[htb]
    \centering
    \includegraphics[trim = 0.0cm 0.0cm 0.0cm 0.0cm, clip=true, width=0.40\textwidth]{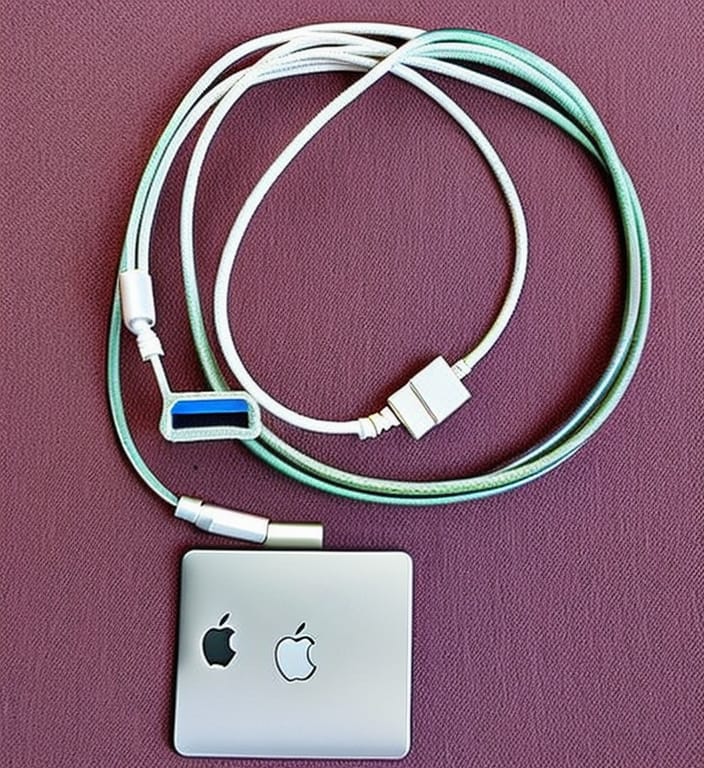} 
    \includegraphics[trim = 0.0cm 0.0cm 0.0cm 0.0cm, clip=true, width=0.40\textwidth]{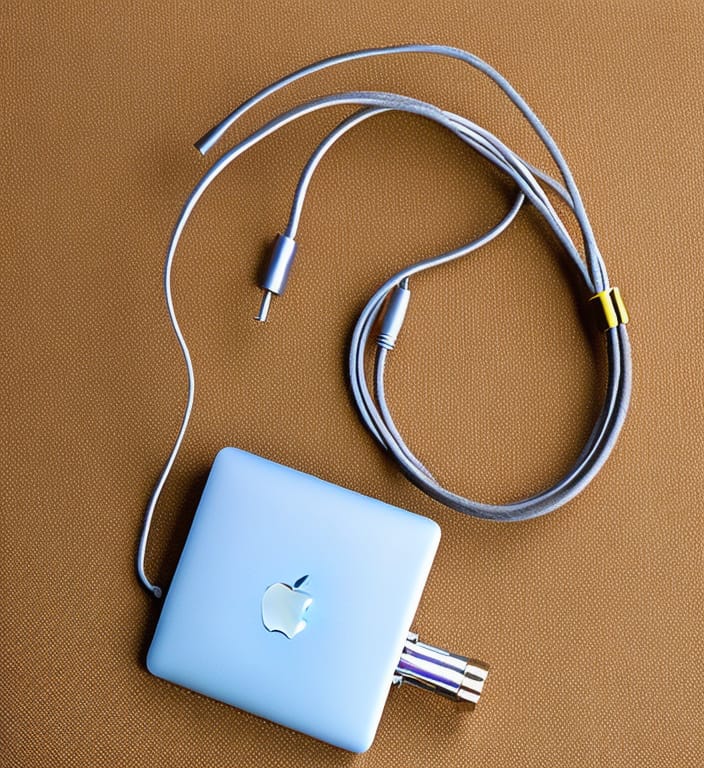} 
    \caption{Two simulated tangled MacBook cable samples, as depicted the Stable Diffusion, a deep learning, text-to-image model released in 2022 that is used to generate images conditioned on text descriptions, most often for recreation, though we choose not to elaborate what that means. The authors include these images so that we can mention Machine Learning in the paper, and increase our odds of getting future funding by an order of magnitude. Note: even though the machine learning-generated cables barely resemble actual photos, the cables are still tangled.}
    \label{fig:RealCables}
\end{figure}


\section{Summary and Conclusions} \label{sec:discussion}
In this study of the macroscopic dynamics of entangled
3+1-dimensional systems, we have seen that the most probable state for MacBook cable inside a backpack is the tangled one, with a rough probabilty of 75.3\% and no effort yet made to estimate the error on the calculation, in large part due to procrastination and lack of time. The result is somewhat satisfying, in that it appears to confirm The Entangled Laptop Cable Problem as valid ans worthy of further study. While the authors admit that they could simply pack their cables in an orderly fashion, it is once again worth noting that some modern macbook chargers no longer contain prongs to wrap cable around.

\section*{Acknowledgments}
We thank our wife for allowing us the free time to write this paper when we should have been doing our taxes, and our cats Jiffy and Nyx for not attacking and destroying the charging cable during its examination, as a new cable can easily cost as much as several takeout order of Thai food. We also acknowledge our university, the local students, and our colleagues for not pestering under during the handful of days while we panicked and finally wrote this draft. Finally, we thank the plethora of Python developers, Stack Overflow contributors, Wikipedia authors, and others who have so generously contributed their time to making numerical analysis easier for the world of academia, and in particular the physics and astronomy students we are most familiar with. 

\newpage
\bibliographystyle{JHEP}
\bibliography{bibliography}

\end{document}